\documentclass[aps,twocolumn,longbibliography]{revtex4-1}
\pdfoutput=1

\usepackage{amsmath,amssymb,mathtools,bm,color,graphicx,empheq}
\usepackage[normalem]{ulem}

\newcommand{\real}{\operatorname{Re}}

\newcommand{\bra}[1]{\langle #1|}
\newcommand{\ket}[1]{|#1\rangle}

\begin{document}

\title{Simulating quantum backflow on a quantum computer}

\author{Arseni Goussev}
\email[Corresponding author. Email: ]{arseni.goussev@port.ac.uk}
\author{Jaewoo Joo}
\affiliation{School of Mathematics and Physics, University of Portsmouth, Portsmouth PO1 3HF, United Kingdom}

\date{\today}

\begin{abstract}
	Quantum backflow is a counterintuitive effect in which the probability density of a free particle moves in the direction opposite to the particle's momentum. If the particle is electrically charged, then the effect can be viewed as the contrast between the direction of electric current and that of the momentum. To date, there has been no direct experimental observation of quantum backflow. However, the effect has been simulated numerically (using classical computers) and optically (using classical light). In this study, we present the first simulation of quantum backflow using a real quantum computer.
\end{abstract}

\maketitle

\section{Introduction}
\label{sec:intro}

\begin{figure}[h]
	\centering
	\includegraphics[width=0.35\textwidth]{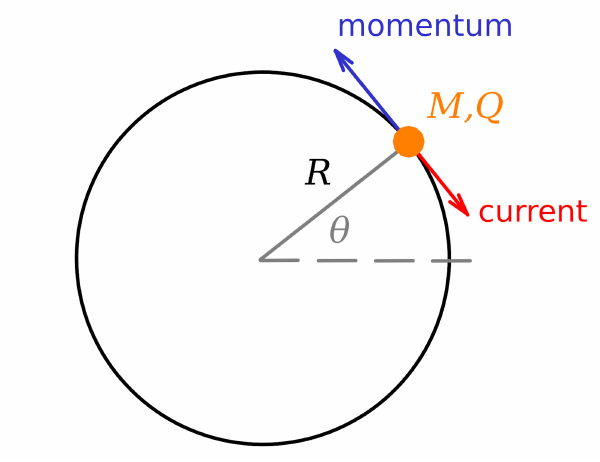}
	\caption{A particular manifestation of the quantum backflow effect.}
	\label{fig1}
\end{figure}

Imagine a bead of mass $M$ and electric charge $Q$ constrained to move without friction along a rigid ring of radius $R$ (Fig.~\ref{fig1}).
Suppose the bead moves freely (i.e., in the absence of any external forces other than the constraint force) in the counterclockwise direction, so that the time derivative of the azimuthal angle $\theta$ is non-negative, $d\theta / dt \ge 0$. The laws of classical physics guarantee that, at every point of the ring, momentum density and electric current density are both nonnegative, i.e., both are either zero or point tangentially to the ring in the direction of increasing $\theta$. Interestingly, the situation can be drastically different if the motion of the bead is governed by the laws of quantum mechanics: There are quantum states of the bead for which its instantaneous momentum and electric current point in the opposite directions (Fig.~\ref{fig1}). This counterintuitive scenario is one particular manifestation of a broad class of phenomena concerned with the classically forbidden flow of probability commonly referred to as {\it quantum backflow} (QB).

QB was first mentioned in the context of the arrival time problem in quantum mechanics \cite{All69time-c, Kij74time}. The first in-depth analysis of QB for a free particle on a line was carried out by Bracken and Melloy \cite{BM94Probability}. In particular, they showed that the effect is weak: Only a small amount of probability -- less than 4$\%$ \footnote{The numerical estimate of this bound, commonly referred to as the Bracken-Melloy constant, has later been improved \cite{EFV05Quantum,PGKW06new} and currently stands at 0.0384517. Recently, the first analytical bound has been reported in Ref.~\cite{TLN23Quantum}, according to which the value of the Bracken-Melloy constant lies between 0.0315 and 0.0725.} -- can be transported in the direction opposite to the particle's momentum. Interestingly, QB can be significantly more pronounced for rotational motion. Thus, the classically forbidden probability transfer can reach values up to approximately 0.116816 in the case of a particle rotating on a ring (Fig.~\ref{fig1})  \cite{Gou21Quantum} and can be arbitrarily high in two-dimensional systems \cite{Str12Large, PPR20Angular, BGS23Unbounded}. Recently, the problem of QB on a ring has also been considered for the case of a massless Dirac fermion \cite{BPPR23Quantum}. The literature on QB is substantial, and reviewing it goes beyond the scope of the present paper. The reader is referred to Ref.~\cite{YH13introduction} for an elementary introduction and to Refs.~\cite{Bra21Probability, BPPR23Quantum} for an extensive list of references to more recent results in the area.

As of today, QB has not been observed experimentally. A promising experimental scheme that could lead to the observation of QB in Bose-Einstein condensates was proposed in Ref.~\cite{PTMM13Detecting}, but, to our knowledge, has never been realized in practice.

While a direct experimental observation of QB remains an open challenge, there has been exciting progress in simulating the QB effect using classical light \cite{EZB20Observation, DGGL22Demonstrating, GDGL23}. The simulations utilize the analogy between the dynamics of quantum particles and the transverse spreading of light beams. In this analogy, the momentum and probability current of a quantum particle are represented by the transverse wave vector and Poynting vector of the light beam, respectively. Thus, the optical experiments reported in Refs.~\cite{EZB20Observation, DGGL22Demonstrating} can be regarded as simulating QB for a particle on a line, whereas the experiment in Ref.~\cite{GDGL23} is the optics counterpart of QB for a two-dimensional rotational motion.

In this paper, we demonstrate how QB for a particle on a ring (Fig.~\ref{fig1}) can be simulated using a quantum computer. The demonstration involves the following two steps: (i) we use $N$ qubits to encode a particle state $| \psi \rangle$ comprised of $2^N$ eigenstates with non-negative angular momentum, and (ii) pass $| \psi \rangle$ through a quantum circuit designed to compute the probability (or electric) current at a given point on the ring. A negative readout is a manifestation of QB. Theoretical results derived in this paper are applicable to the case of arbitrary $N$, whereas the actual experimental demonstration, utilizing the IBM-Q quantum computer \cite{IBMQ}, is performed for the cases of $N=1$ and $N=2$.

The paper is organized as follows. In Sec.~\ref{sec:QB}, we specify the system and formulate the QB effect. In Sec.~\ref{sec:current}, we introduce a finite-dimensional probability current operator and derive its decomposition in terms of sums of tensor products of one-qubit operators. This decomposition is what makes the following quantum-computer simulation of QB possible. In Sec.~\ref{sec:example}, we construct a concrete example, valid for arbitrary $N$, of a quantum state exhibiting QB. We later use this state in our quantum simulation. Section~\ref{sec:experiment} presents an experimental simulation of QB performed on the IBM-Q quantum computer. In Sec.~\ref{sec:conclusion} we summarize our work and make concluding remarks. Throughout the paper, we set the particle mass and electric charge, the ring radius, and the Planck constant equal to unity, i.e. $M = Q = R = \hbar = 1$.

\section{Quantum backflow for circular motion}
\label{sec:QB}

The Hamiltonian of the particle-on-a-ring system (Fig.~\ref{fig1}) is $\hat{H} = \frac{1}{2} \hat{L}^2$, where $\hat{L} = -i \frac{d}{d \theta}$ is the angular momentum operator. Stationary states $| m \rangle$ that simultaneously diagonalize $\hat{H}$ and $\hat{L}$, namely
\begin{equation*}
	\hat{H} | m \rangle  = \frac{m^2}{2} | m \rangle \,, \qquad \hat{L} | m \rangle  = m | m \rangle \,,
\end{equation*}
are given by
\begin{equation}
	\langle \theta | m \rangle = \frac{e^{i m \theta}}{\sqrt{2 \pi}} \qquad (m \in \mathbb{Z}) \,.
\label{eigenstates}
\end{equation}
The states are orthonormal,
\begin{equation*}
	\langle m | n \rangle = \int_{-\pi}^{\pi} d\theta \, \langle m | \theta \rangle \langle \theta | n \rangle = \delta_{mn} \,,
\end{equation*}
and form a complete basis.

Now suppose that the particle is in a state $| \psi \rangle$ given by a superposition of $2^N$ stationary states of the lowest possible energy and non-negative angular momentum:
\begin{equation}
	| \psi \rangle = \sum_{m = 0}^{2^N - 1} a_m | m \rangle \,.
\label{state}
\end{equation}
The expansion coefficients $a_m$ satisfy the normalization condition,
\begin{equation}
	\sum_{m = 0}^{2^N - 1} |a_m|^2 = 1 \,.
\label{normalization}
\end{equation}
By construction, any angular momentum measurement performed on $| \psi \rangle$ is guaranteed to return a non-negative result. However, the probability (or electric) current
\begin{equation}
	J = \left. \real \left\{ \langle \psi | \theta \rangle \left( -i \frac{d}{d \theta} \right) \langle \theta | \psi \rangle \right\} \right|_{\theta = \theta_0}
\label{current-1}
\end{equation}
at some fixed point $\theta = \theta_0$ on the ring {\it can} be negative. This is the essence of the QB effect.

\section{Current operator}
\label{sec:current}

We now introduce a probability current operator and derive its representation in terms of tensor products of one-qubit gates. Without any loss of generality, and in order to simplify the calculations to follow, we set $\theta_0 = 0$. Then, substituting Eq.~\eqref{state} into Eq.~\eqref{current-1} and making use of  and Eq.~\eqref{eigenstates}, we obtain
\begin{align}
	J &= \frac{1}{2 \pi} \real \sum_{m,n = 0}^{2^N-1} a_m^* n a_n \nonumber \\
	&= \frac{1}{4 \pi} \sum_{m,n = 0}^{2^N-1} a_m^* (m + n) a_n \,.
\label{J_in_terms_of_a_m}
\end{align}
Alternatively, $J$ can be written as
\begin{equation*}
	J = \frac{1}{4 \pi} \langle \psi | \hat{\mathcal{J}}_N | \psi \rangle \,,
\end{equation*}
where the operator
\begin{equation}
	\hat{\mathcal{J}}_N = \sum_{m,n = 0}^{2^N-1} | m \rangle (m + n) \langle n |
\label{J_N_def}
\end{equation}
represents the (scaled) probability current at $\theta_0 = 0$ for quantum states in the $2^N$-dimensional subspace of the Hilbert space spanned by $| 0 \rangle$, $| 1 \rangle$, $\ldots$, $| 2^N-1 \rangle$.

In the rest of this section, we show how the operator $\hat{\mathcal{J}}_N$, for any $N = 1, 2, 3, \ldots$, can be decomposed into a sum of tensor products of the following two-dimensional operators: the identity operator $\hat{I}$ and the Pauli gates $\hat{X}$ and $\hat{Z}$. The main result of this section is given by Eq.~\eqref{decomposition_of_J_N}. This decomposition is essential for one's ability to simulate QB on a quantum computer.

\subsection{$N=1$ case}
\label{sec:theory_N_1}

We begin by considering the $N=1$ case. In the matrix representation defined by
\begin{equation*}
	| 0 \rangle =
	\begin{pmatrix}
		1 \\ 0
	\end{pmatrix} \,, \qquad | 1 \rangle =
	\begin{pmatrix}
		0 \\ 1
	\end{pmatrix} \,,
\end{equation*}
the current operator reads
\begin{equation}
	\hat{\mathcal{J}}_1 =
	\begin{pmatrix}
		0 & 1 \\
		1 & 2
	\end{pmatrix} \,.
\label{J_1_matrix}
\end{equation}
Clearly, $\hat{\mathcal{J}}_1$ can be decomposed as
\begin{equation}
	\hat{\mathcal{J}}_1 = \hat{I} + \hat{X} - \hat{Z} \,,
\label{decomposition_N_1}
\end{equation}
where
\begin{equation*}
	\hat{I} =
	\begin{pmatrix}
		1 & 0 \\
		0 & 1
	\end{pmatrix} \,, \quad \hat{X} =
	\begin{pmatrix}
		0 & 1 \\
		1 & 0
	\end{pmatrix} \,, \quad \hat{Z} =
	\begin{pmatrix}
		1 & 0 \\
		0 & -1
	\end{pmatrix} \,.
\end{equation*}

\subsection{$N=2$ case}
\label{sec:theory_N_2}

We now turn to the $N=2$ case. Writing
\begin{equation*}
	| 0 \rangle = 
	\begin{pmatrix}
		1 \\ 0 \\ 0 \\ 0
	\end{pmatrix} , \quad
	| 1 \rangle = 
	\begin{pmatrix}
		0 \\ 1 \\ 0 \\ 0
	\end{pmatrix} , \quad
	| 2 \rangle = 
	\begin{pmatrix}
		0 \\ 0 \\ 1 \\ 0
	\end{pmatrix} , \quad
		| 3 \rangle = 
	\begin{pmatrix}
		0 \\ 0 \\ 0 \\ 1
	\end{pmatrix} ,
\end{equation*}
we have
\begin{equation*}
	\hat{\mathcal{J}}_2 =
	\begin{pmatrix}
		0 & 1 & 2 & 3 \\
		1 & 2 & 3 & 4 \\
		2 & 3 & 4 & 5 \\
		3 & 4 & 5 & 6
	\end{pmatrix} \,.
\end{equation*}
The last matrix can be rewritten as follows:
\begin{equation}
	\hat{\mathcal{J}}_2 =
	\begin{pmatrix}
		\hat{\mathcal{J}}_1 & \hat{\mathcal{J}}_1 + 2 \hat{C}_1 \\
		\hat{\mathcal{J}}_1 + 2 \hat{C}_1 & \hat{\mathcal{J}}_1 + 4 \hat{C}_1
	\end{pmatrix} \,,
\label{J_2_as_matrix_of_J_1}
\end{equation}
where $\hat{\mathcal{J}}_1$ is given by Eqs.~\eqref{J_1_matrix} and \eqref{decomposition_N_1}, and
\begin{equation*}
	\hat{C}_1 =
	\begin{pmatrix}
		1 & 1 \\ 1 & 1
	\end{pmatrix} \,.
\end{equation*}
Then,
\begin{align}
	\hat{\mathcal{J}}_2 &=
	\begin{pmatrix}
		1 & 1 \\
		1 & 1
	\end{pmatrix} \otimes \hat{\mathcal{J}}_1 +
	\begin{pmatrix}
		0 & 2 \\
		2 & 4
	\end{pmatrix} \otimes \hat{C}_1 \nonumber \\
	&= \hat{C}_1 \otimes \hat{\mathcal{J}}_1 + 2 \hat{\mathcal{J}}_1 \otimes \hat{C}_1 \,.
\label{towards_decomposition_N_2}
\end{align}
Noticing that
\begin{equation}
	\hat{\mathcal{J}}_1 = \hat{C}_1 - \hat{Z} \,,
	\label{J_1_in_terms_of_C_1}
\end{equation}
we find
\begin{equation}
	\hat{\mathcal{J}}_2 = 3 \hat{C}_1 \otimes \hat{C}_1 - \hat{C}_1 \otimes \hat{Z} - 2 \hat{Z} \otimes \hat{C}_1 \,.
	\label{J_2_in_terms_of_Z}
\end{equation}
Finally, using
\begin{equation}
	\hat{C}_1 = \hat{I} + \hat{X} \,,
\label{C_1_in_terms_of_X}
\end{equation}
we arrive at the following decomposition:
\begin{equation}
	\hat{\mathcal{J}}_2 = 3 (\hat{I} + \hat{X}) \otimes (\hat{I} + \hat{X}) - (\hat{I} + \hat{X}) \otimes \hat{Z} - 2 \hat{Z} \otimes (\hat{I} + \hat{X}) \,.
\label{decomposition_N_2}
\end{equation}

\subsection{General case}
\label{sec:theory_general}

We now generalize the method of Sec.~\ref{sec:theory_N_2} to construct an explicit decomposition of $\hat{\mathcal{J}}_N$, for arbitrary $N$. To this end, we first write $\hat{J}_N$, defined by Eq.~\eqref{J_N_def}, in the matrix form. Following the convention adopted in Secs.~\ref{sec:theory_N_1} and \ref{sec:theory_N_2}, we take $| m \rangle$ to be represented by the $2^N$-dimensional column vector with the $n^{\text{th}}$ element equal to $\delta_{mn}$. Then,
\begin{equation*}
	\hat{\mathcal{J}}_N =
	\begin{pmatrix}
		0 & 1 & 2 & \cdots & 2^N-1 \\
		1 & 2 & 3 & \cdots & 2^N \\
		2 & 3 & 4 & \cdots & 2^N+1 \\
		\cdots & \cdots & \cdots & \cdots & \cdots \\
		2^N-1 & 2^N & 2^N+1 & \cdots & 2^{N+1}-2
	\end{pmatrix} \,.
\end{equation*}
Since the last matrix can be written as
\begin{widetext}
\begin{equation*}
	\left(
	\begin{array}{cccc|cccc}
		0 & 1 &  \cdots & 2^{N-1} - 1 & 0 + 2^{N-1} & 1 + 2^{N-1} & \cdots &  (2^{N-1} - 1) + 2^{N-1} \\
		1 & 2 & \cdots & 2^{N-1} & 1 + 2^{N-1} & 2 + 2^{N-1} & \cdots & 2^{N-1} + 2^{N-1} \\
		\cdots & \cdots & \cdots & \cdots & \cdots & \cdots & \cdots & \cdots \\
		2^{N-1} - 1 & 2^{N-1} & \cdots & 2^N - 2 & (2^{N-1} - 1) + 2^{N-1} & 2^{N-1} + 2^{N-1} & \cdots & (2^N - 2) + 2^{N-1} \\
		\hline
		0 + 2^{N-1} & 1 + 2^{N-1} & \cdots & (2^{N-1} - 1) + 2^{N-1} & 0 + 2^N & 1 + 2^N & \cdots & (2^{N-1} - 1) + 2^N \\
		1 + 2^{N-1} & 2 + 2^{N-1} & \cdots & 2^{N-1} + 2^{N-1} & 1 + 2^N & 2 + 2^N & \cdots & 2^{N-1} + 2^N \\
		\cdots & \cdots & \cdots & \cdots & \cdots & \cdots & \cdots & \cdots \\
		(2^{N-1} - 1) + 2^{N-1} & 2^{N-1} + 2^{N-1} & \cdots & (2^N - 2) + 2^{N-1} & (2^{N-1} - 1) + 2^N & 2^{N-1} + 2^N & \cdots & (2^N - 2) + 2^N
	\end{array}
	\right) \,,
\end{equation*}
\end{widetext}
it is easy to see that [cf.~Eq.~\eqref{J_2_as_matrix_of_J_1}]
\begin{equation}
	\hat{\mathcal{J}}_N =
	\begin{pmatrix}
		\hat{\mathcal{J}}_{N-1} \; & \hat{\mathcal{J}}_{N-1} + 2^{N-1} \hat{C}_{N-1} \\[0.2cm]
		 \hat{\mathcal{J}}_{N-1} + 2^{N-1} \hat{C}_{N-1} \; & \hat{\mathcal{J}}_{N-1} + 2^N \hat{C}_{N-1}
	\end{pmatrix} \,,
\label{J_N+1_as_matrix_of_J_N}
\end{equation}
where $\hat{C}_{N-1}$ is the $2^{N-1} \times 2^{N-1}$ matrix of ones. It follows from Eq.~\eqref{J_N+1_as_matrix_of_J_N} that
\begin{equation}
	\hat{\mathcal{J}}_N = \hat{C}_1 \otimes \hat{\mathcal{J}}_{N-1} + 2^{N-1} \hat{\mathcal{J}}_1 \otimes \hat{C}_{N-1} \,.
\label{towards_decomposition_N+1}
\end{equation}
Substituting
\begin{equation*}
	\hat{C}_{N-1} = \underbrace{ \hat{C}_1 \otimes \hat{C}_1 \otimes \cdots \otimes \hat{C}_1 }_{N-1 \text{ times}} = \hat{C}_1^{\otimes (N-1)}
\end{equation*}
into Eq.~\eqref{towards_decomposition_N+1}, we get the following recurrence relation [cf.~Eq.~\eqref{towards_decomposition_N_2}]:
\begin{equation*}
	\hat{\mathcal{J}}_N = \hat{C}_1 \otimes \hat{\mathcal{J}}_{N-1} + 2^{N-1} \hat{\mathcal{J}}_1 \otimes \hat{C}_1^{\otimes (N-1)} \,.
\end{equation*}
It is straightforward to verify that the solution to this recurrence relation is [cf.~Eq.~\eqref{towards_decomposition_N_2}]
\begin{equation*}
	\hat{\mathcal{J}}_N = \sum_{n=0}^{N-1} 2^n \hat{C}_1^{\otimes (N-1-n)} \otimes \hat{\mathcal{J}}_1 \otimes \hat{C}_1^{\otimes n} \,.
\end{equation*}
Substituting Eq.~\eqref{J_1_in_terms_of_C_1} into the last expression, and using $\sum_{n=0}^{N-1} 2^n = 2^N - 1$, we obtain [cf.~Eq.~\eqref{J_2_in_terms_of_Z}]
\begin{equation*}
	\hat{\mathcal{J}}_N = (2^N - 1) \hat{C}_1^{\otimes N} - \sum_{n=0}^{N-1} 2^n \hat{C}_1^{\otimes (N-1-n)} \otimes \hat{Z} \otimes \hat{C}_1^{\otimes n} \,.
\end{equation*}
Finally, using Eq.~\eqref{C_1_in_terms_of_X}, we arrive at the following explicit decomposition of $\hat{\mathcal{J}}_N$ [cf.~Eq.~\eqref{decomposition_N_2}]:
\begin{equation}
	\begin{split}
		\MoveEqLeft
		\hat{\mathcal{J}}_N = (2^N - 1) (\hat{I} + \hat{X})^{\otimes N} \\
		& - \sum_{n=0}^{N-1} 2^n (\hat{I} + \hat{X})^{\otimes (N-1-n)} \otimes \hat{Z} \otimes (\hat{I} + \hat{X})^{\otimes n} \,.
	\end{split}
\label{decomposition_of_J_N}
\end{equation}
Equation~\eqref{decomposition_of_J_N} constitutes the main result of this section.

\section{Example of a backflowing state}
\label{sec:example}

Decomposition~\eqref{decomposition_of_J_N} allows us to devise a quantum computing circuit for measuring the probability current at a fixed point on the ring. What we also need for a quantum simulation of QB is a quantum state that would give rise to a substantially negative probability current. In this section, we present an explicit example of such a state.

Consider the state $| \psi \rangle$ defined by Eq.~\eqref{state} with
\begin{equation}
	a_m = \frac{3 m - 2 (2^N - 1) }{\sqrt{(2^{N+1}+1) (2^N - 1) 2^{N-1}}} \,.
\label{example_state}
\end{equation}
The state is normalized. Indeed, using identities $\sum_{m=0}^{2^N-1} m = (2^N-1) 2^{N-1}$ and $\sum_{m=0}^{2^N-1} m^2 = \frac{1}{3} (2^{N+1}-1) (2^N - 1) 2^{N-1} $, it is straightforward to verify that the normalization condition~\eqref{normalization} is fulfilled.

Now, substituting Eq.~\eqref{example_state} into Eq.~\eqref{J_in_terms_of_a_m} and performing a straightforward calculation, taking into account that $\sum_{m=0}^{2^N-1} m^3 = 2^{2N-2} (2^N-1)^2$, we find
\begin{equation*}
	J = -\frac{1}{4 \pi} \frac{2^N (2^N-1)}{2^{N+1}+1} \,.
\end{equation*}
Clearly $J < 0$, for all $N \ge 1$, meaning that $| \psi \rangle$, defined by Eq.~\eqref{example_state} exhibits QB. In particular,
\begin{equation}
	J = -\frac{1}{10 \pi} \simeq -0.031831 \quad \text{for} \quad N = 1
\label{J_theor_N_1}
\end{equation}
and
\begin{equation}
	J = -\frac{3}{9 \pi} \simeq -0.106103 \quad \text{for} \quad N = 2 \,.
\label{J_theor_N_2}
\end{equation}

For the purpose of clarity, we would like to point out that the state defined by Eq.~\eqref{example_state} is not the only state exhibiting negative probability current (see Ref.~\cite{Gou21Quantum} for other examples of backflowing states.) Nor is it the state maximizing the backflow probability transfer \cite{Gou21Quantum}. The main reasons we use the state given by Eq.~\eqref{example_state} in our study are its simplicity -- specifically, the fact that the expansion coefficients have a simple linear dependence on the quantum number $m$ -- and its generality, as the state gives rise to negative probability current for any $N$.

It is also interesting to note that, for the state defined by Eq.~\eqref{example_state}, $J \to -\infty$ as $N \to \infty$. This example shows that, just as in the particle-on-a-line case~\cite{BM94Probability}, the instantaneous probability current for non-negative angular momentum states in a ring is unbounded from below.

\section{Implementation of quantum backflow on a quantum computer}
\label{sec:experiment}

Equipped with the decomposition~\eqref{decomposition_of_J_N} and the explicit backflowing state example~\eqref{example_state}, we proceed to simulating QB on the IBM-Q quantum computer \cite{IBMQ}.
\begin{figure}[h]
	\centering
	\includegraphics[width=0.4\textwidth]{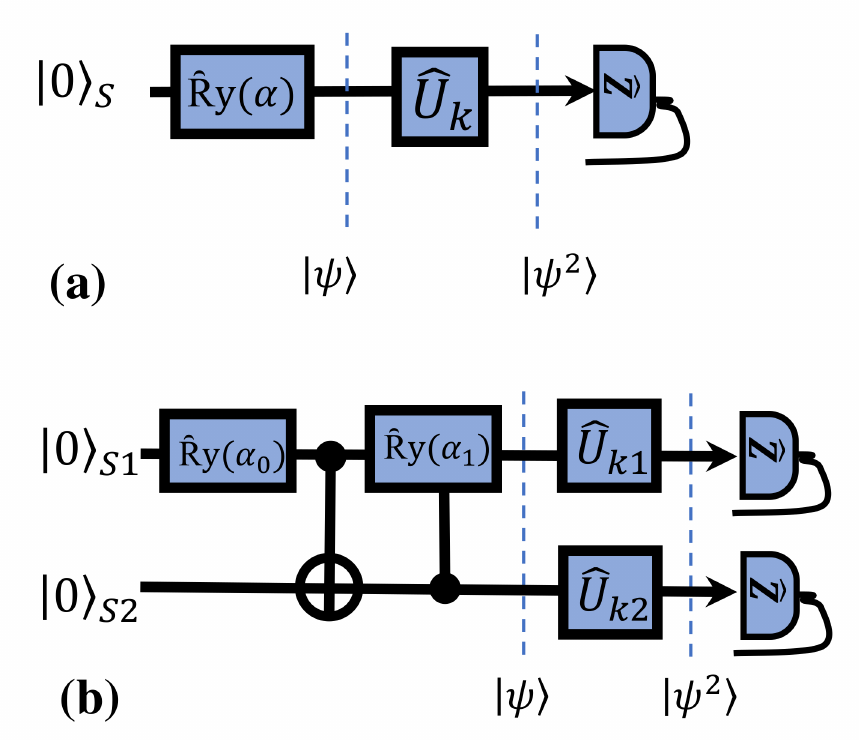}
	\caption{Quantum circuits for computing $\langle \hat{\cal{V}}_k \rangle \equiv \bra{\psi}\hat{\cal V}_k \ket{\psi}$ for (a) $N=1$ and (b) $N=2$.}
	\label{fig2}
\end{figure}
The quantum circuits used in our simulation are schematically shown in Figs.~\ref{fig2}(a) and \ref{fig2}(b) for the cases of $N=1$ and $N=2$, respectively. The angular momentum states $| m \rangle$, with $m = 0, \ldots, 2^N-1$, of the particle on the ring are represented by system qubits that are denoted by $S$ in the $N=1$ case, and by $S_1$ and $S_2$ in the $N=2$ case. More specifically, we have
\begin{equation*}
	| 0 \rangle = | 0 \rangle_S \quad \text{and} \quad | 1 \rangle = | 1 \rangle_S
\end{equation*}
for $N=1$, and
\begin{align}
\begin{split}
	&| 0 \rangle = | 0 \rangle_{S_1} | 0 \rangle_{S_2} = |00\rangle \\
	&| 1 \rangle = | 0 \rangle_{S_1} | 1 \rangle_{S_2} = |01\rangle \\
	&| 2 \rangle = | 1 \rangle_{S_1} | 0 \rangle_{S_2} = |10\rangle \\
	&| 3 \rangle = | 1 \rangle_{S_1} | 1 \rangle_{S_2} = |11\rangle
\end{split}
\label{S1_S2_def}
\end{align}
for $N=2$.

The simulation consists of the following four stages. First, we prepare the system qubit(s) in state $|0\rangle_S$ for $N=1$, and in states $|0\rangle_{S_1}$ and $|0\rangle_{S_2}$ for $N=2$. Second, we apply a set of gates to the system qubit(s) to create the target input state $\ket{\psi}$, Eq.~\eqref{state}, with the expansion coefficients given by Eq.~\eqref{example_state}. Third, we act with single-qubit gates $\hat{U}_{k}$ for $N=1$, and with $\hat{U}_{k1}$ and $\hat{U}_{k2}$ for $N=2$, on $\ket{\psi}$ to obtain a new state $\ket{\psi^2}$. The gates $\hat{U}_k$, $\hat{U}_{k1}$, and $\hat{U}_{k2}$ are to be defined below. Finally, the qubits of $\ket{\psi^2}$ are measured in the $\hat{Z}$ basis and the measurement outcome probabilities are used to compute the expectation value of operators $\hat{\cal V}_k$. As we explain in detail below, the operators $\hat{\cal V}_k$ are elementary $N$-qubit gates allowing one to represent the current operator as
\begin{equation}
	\hat{\cal J}_N  = \lambda_0 \hat{I}^{\otimes N} + \sum_{k=1}^{K_N} \lambda_k \hat{\cal V}_k
\label{J_N_in_terms_of_Vs}
\end{equation}
with $K_1 = 2$, $K_2 = 7$, and $\lambda_k$'s being some real numbers.

 We now provide further details of the outlined simulation procedure and present the experimental results, separately in the $N = 1$ and $N = 2$ case.

\subsection{$N = 1$ case}

Here we detail the experimental procedure, schematically illustrated in Fig.~\ref{fig2}(a). Hereinafter, we use the notation $\langle \cdot \rangle = \langle \psi | \cdot | \psi \rangle$, with $| \psi \rangle$ given by Eqs.~\eqref{state} and \eqref{example_state}.

Let us rewrite Eq.~\eqref{decomposition_N_1} as $\hat{\cal J}_1 = \hat{I} + \hat{\cal{V}}_1 - \hat{\cal{V}}_2$ with $\hat{\cal{V}}_1 = \hat{X}$ and $\hat{\cal{V}}_2 = \hat{Z}$. This representation corresponds to Eq.~\eqref{J_N_in_terms_of_Vs} with $N = 1$, $K_1 = 2$, $\lambda_0 = \lambda_1 = 1$ and $\lambda_2 = -1$. Consequently, we have
\begin{equation}
	\langle \hat{\cal J}_1 \rangle = 1 + \langle \hat{\cal{V}}_1 \rangle - \langle \hat{\cal{V}}_2 \rangle \,.
\label{J_1_in_terms_of_Vk}
\end{equation}
The following procedure allows us to experimentally measure $\langle \hat{\cal{V}}_1 \rangle$ and $\langle \hat{\cal{V}}_2 \rangle$.

We first initialize the system qubit, $S$, in state $\ket{0}_S$. We then transform $\ket{0}_S$ into the desired backflowing state
\begin{align*}
	\ket{\psi} &= \frac{1}{\sqrt{5}} \left( -2 \ket{0}_S + \ket{1}_S \right) \\ &\simeq -0.894427 \ket{0}_S + 0.447214 \ket{1}_S \,,
\end{align*}
whose expansion coefficients are obtained from Eq.~\eqref{example_state} by setting $N = 1$ and $m = 0, 1$. The transformation is achieved by acting on $\ket{0}_S$ with the $Y$-rotation operator $\hat{R}_y(\alpha) \equiv \exp \left( -i \frac{\alpha}{2} \hat{Y} \right)$:
\begin{equation*}
	\ket{\psi} = \hat{R}_y(\alpha) \ket{0}_S
\end{equation*}
with $\alpha \simeq 5.35589$. Next, we apply gate $\hat{U}_k$ to the system qubit to obtain
\begin{equation}
	\ket{\psi^2} = \hat{U}_k \ket{\psi} \,.
\label{from_psi2_to_psi3_N_1}
\end{equation}
Here, $\hat{U}_1$ is the Hadamard gate,
\begin{equation*}
	\hat{U}_1 = \hat{H}_d \equiv \frac{1}{\sqrt{2}} (\hat{X} + \hat{Z}) \,, 
\end{equation*}
and $\hat{U}_2$ is the identity operator,
\begin{equation*}
	\hat{U}_2 = \hat{I} \,.
\end{equation*}
This choice of the transformations $\hat{U}_1$ and $\hat{U}_2$ allows us to experimentally determine the sought expectation values $\langle \hat{\cal{V}}_1 \rangle$ and $\langle \hat{\cal{V}}_2 \rangle$ by measuring $\ket{\psi^2}$ in the $\hat{Z}$ basis. This works as follows.

The probabilities that the measurement will collapse $\ket{\psi^2}$ onto $\ket{0}_S$ and $\ket{1}_S$ are given by
\begin{align*}
	&P_0 = \bra{\psi^2} 0 \rangle_S \langle {0} \ket{\psi^2} \,, \\
	&P_1 = \bra{\psi^2} 1 \rangle_{S} \langle {1} \ket{\psi^2} \,,
\end{align*}
respectively, and so
\begin{equation*}
	P_0 - P_1 = \bra{\psi^2} \hat{Z} \ket{\psi^2} \,.
\end{equation*}
For $k = 1$, we have
\begin{equation*}
	\ket{\psi^2} = \hat{U}_1 \ket{\psi} = \hat{H}_d \ket{\psi} \,,
\end{equation*}
which means that
\begin{equation*}
	P_0 - P_1 = \bra{\psi} \hat{H}_d \hat{Z} \hat{H}_d \ket{\psi} = \bra{\psi} \hat{X} \ket{\psi} = \langle \hat{\cal{V}}_1 \rangle \,,
\end{equation*}
where we have used the identity $\hat{X} = \hat{H}_d \hat{Z} \hat{H}_d$. For $k = 2$, we have
\begin{equation*}
	\ket{\psi^2} = \hat{U}_2 \ket{\psi} = \ket{\psi} \,,
\end{equation*}
implying that
\begin{equation*}
	P_0 - P_1 = \bra{\psi} \hat{Z} \ket{\psi} = \langle \hat{\cal{V}}_2 \rangle \,.
\end{equation*}
This is how, by measuring the difference $P_0 - P_1$ in the quantum circuit shown in Fig.~\ref{fig2}(a), we can evaluate the expectation values $\langle \hat{\cal{V}}_k \rangle$ experimentally.

We used the IBM-Q machine named $ibmq_{-}osaka$ \cite{IBMQ} to perform 8000 projective measurements for each $\langle \hat{\cal{V}}_k \rangle$. The measurement results are $P_0 \simeq 0.095757$ and $P_1 \simeq 0.904243$ for $k = 1$, implying $\langle \hat{\cal V}_1 \rangle \simeq -0.808487$, and $P_0 \simeq 0.793382$, $P_1 \simeq 0.206618$ for $k = 2$, implying $\langle \hat{\cal V}_2 \rangle \simeq 0.586764$. Thus, in view of Eq.~\eqref{J_1_in_terms_of_Vk}, the experimentally obtained value of probability current in the $N = 1$ case is
\begin{equation*}
	J = {1 \over 4 \pi} \langle \hat{\cal J}_1 \rangle \simeq -0.031453 \,.
\end{equation*}
The experimental value is negative, signifying QB, and is reasonably close to the theoretical one, given by Eq.~\eqref{J_theor_N_1}. The relative error is approximately $2.2\%$.

\subsection{$N = 2$ case}

The quantum circuit used for simulating QB in the $N=2$ case is shown in Fig.~\ref{fig2}(b). It utilizes two system qubits, $S_1$ and $S_2$, and seven two-qubit operators $\hat{\cal{V}}_k$, with $k = 1, \ldots, 7$. According to  Eq.~\eqref{decomposition_N_2}, the (scaled) probability current operator $\hat{\cal J}_2$ can be represented by Eq.~\eqref{J_N_in_terms_of_Vs} with $N = 2$, $K_2 = 7$,  $\hat{\cal V}_1 = \hat{I} \otimes \hat{X}$, $\hat{\cal V}_2= \hat{X}\otimes \hat{I}$,  $\hat{\cal V}_3 = \hat{Z}\otimes \hat{I}$, $\hat{\cal V}_4 = \hat{I}\otimes \hat{Z}$, $\hat{\cal V}_5 = \hat{X}\otimes \hat{X}$  $\hat{\cal V}_6 = \hat{Z}\otimes \hat{X}$, $\hat{\cal V}_7 = \hat{X}\otimes  \hat{Z}$, and $\lambda_0 = \lambda_1 = \lambda_2 = \lambda_5 = 3$, $\lambda_3 = \lambda_6 = -2$, $\lambda_4 = \lambda_7 = -1$. Thus, the expectation value of $\hat{\cal J}_2$ reads
\begin{equation}
	\langle \hat{\cal J}_2 \rangle = 3 + 3\langle \hat{\cal V}_1 \rangle + 3 \langle \hat{\cal V}_2 \rangle - 2 \langle \hat{\cal V}_3 \rangle - \langle \hat{\cal V}_4 \rangle + 3 \langle \hat{\cal V}_5 \rangle - 2 \langle \hat{\cal V}_6 \rangle - \langle \hat{\cal V}_7  \rangle \,.
\label{J_2_in_terms_of_Vk}
\end{equation}
The expectation values $\langle \hat{\cal{V}}_k \rangle$, $k = 1, \ldots, 7$, are measured as follows.

We begin by initializing the system qubits in the $\ket{00}$ state [see Eq.~\eqref{S1_S2_def}]. Then, we apply to $\ket{00}$ a sequence of one- and two-qubit transformations aiming to generate the following (entangled) backflowing state:
\begin{align*}
	\ket{\psi} &= {1 \over \sqrt{6}} (-2\ket{00} - \ket{01} + \ket{11}) \\
	&\simeq -0.816497 \ket{00} -0.408248 \ket{01} + 0.408248 \ket{11} \,.
\end{align*}
The expansion coefficients of $\ket{\psi}$ are obtained from Eq.~\eqref{example_state} by setting $N = 2$ and $m = 0, 1, 2, 3$. In particular, the overlap probabilities between $\ket{\psi}$ and the basis states $\ket{00}$, $\ket{01}$, $\ket{10}$, and $\ket{11}$ are $|\langle 00 | \psi \rangle |^2 \simeq 0.66667$, $|\langle 01 | \psi \rangle |^2 = |\langle 11 | \psi \rangle |^2 \simeq 0.166667$, and $|\langle 10 | \psi \rangle |^2 = 0$, respectively.

Experimentally, the transformation from $\ket{00}$ to $\ket{\psi}$ is performed via successive application of three operators \cite{LJM+20Variational, JM21}: an $\hat{R}_y$ gate, a CNOT gate $CX_{S_1,S_2}$, and a controlled-$\hat{R}_y$ gate. Thus, as illustrated in Fig.~\ref{fig2}(b),
\begin{equation*}
	\ket{\psi } = CR_{S_2,S_1} (\alpha_1) \, CX_{S_1,S_2} \left( \hat{R}_y (\alpha_0) \otimes \hat{I} \right) \ket{00}
\end{equation*}
with $\alpha_0 \simeq 7.51414$ and $\alpha_1 \simeq 4.7124$. In the actual experiment, carried out on $ibmq_{-}osaka$ \cite{IBMQ}, the controlled-rotation gate $CR_{S_2,S_1}(\alpha_1)$ was realized by $\hat{R}_y (\pm \alpha_1/2)$ and two CNOT gates \cite{Nielsen_Chuang}. The accuracy of the experimental state preparation can be characterized by following probabilities of the overlap between $\ket{\psi}$ and the two-qubit basis states: $|\langle 00 | \psi \rangle |^2 \simeq 0.655584$, $|\langle 01 | \psi \rangle |^2 \simeq 0.171429$, $|\langle 10 | \psi \rangle |^2 \simeq 0.002444$, and $|\langle 11 | \psi \rangle |^2 \simeq 0.170543$, which are to be compared against the corresponding theoretical values stated above.  

Once the backflowing state $\ket{\psi}$ has been prepared, the rest of the simulation follows the same steps as in the $N = 1$ case. The transformation from $\ket{\psi}$ to $\ket{\psi^2}$ [see Fig.~\ref{fig2}(b)] is performed by means of two single-qubit gates, $\hat{U}_{k1}$ and $\hat{U}_{k2}$:
\begin{equation*}
	\ket{\psi^2} = \hat{U}_{k1} \otimes \hat{U}_{k2} \ket{\psi}
\end{equation*}
[cf.~Eq.~\eqref{from_psi2_to_psi3_N_1} in the $N = 1$ case]. The single-qubit gate pairs $\{ \hat{U}_{k1}, \hat{U}_{k2} \}$ are $\{\hat{I},  \hat{H}_d \}$ for $k = 1$ and $k = 6$, $\{ \hat{H}_d,\hat{I} \}$ for $k = 2$ and $k = 7$, $\{ \hat{I},\hat{I} \}$ for $k = 3$ and $k = 4$, and $\{ \hat{H}_d, \hat{H}_d \}$ for $k = 5$. For each $k$, $| \psi^2 \rangle$ is measured in the $\hat{Z}$ basis to obtain the probabilities
\begin{align*}
	&P_{00} = \bra{\psi^2} 00 \rangle \langle 00 \ket{\psi^2} \,, \\
	&P_{01} = \bra{\psi^2} 01 \rangle \langle 01 \ket{\psi^2} \,, \\
	&P_{10} = \bra{\psi^2} 10 \rangle \langle 10 \ket{\psi^2} \,, \\
	&P_{11} = \bra{\psi^2} 11 \rangle \langle 11 \ket{\psi^2} \,,
\end{align*}
which are then used to calculate the desired expectation values $\langle \hat{\cal V}_1 \rangle$, $\langle \hat{\cal V}_2 \rangle$, $\ldots$, $\langle \hat{\cal V}_7 \rangle$. For $k=1$,
\begin{align*}
	&P_{00} - P_{01} + P_{10} - P_{11} \\ 
	&\quad = (P_{00} + P_{10}) - (P_{01} + P_{11}) \\
	&\quad = \langle \psi^2 | \Big( \hat{I} \otimes | 0 \rangle_{S_2} \langle 0 | \Big) \psi^2 \rangle - \langle \psi^2 | \Big( \hat{I} \otimes | 1 \rangle_{S_2} \langle 1 | \Big) \psi^2 \rangle \\
	&\quad = \langle \psi^2 | \Big( \hat{I} \otimes \hat{Z} \Big) \psi^2 \rangle \\
	&\quad = \langle \psi | \Big( \hat{I} \otimes \hat{H}_d \Big) \Big( \hat{I} \otimes \hat{Z} \Big) \Big( \hat{I} \otimes \hat{H}_d \Big) | \psi \rangle \\
	&\quad = \langle \psi | \Big( \hat{I} \otimes (\hat{H}_d \hat{Z} \hat{H}_d) \Big) | \psi \rangle \\
	&\quad = \langle \psi | \hat{I} \otimes \hat{X} | \psi \rangle \\
	&\quad = \langle \hat{\cal V}_1 \rangle \,.
\end{align*}
The other expectation values are obtained in a similar way. The resulting expressions are as follows:
\begin{align*}
	&\text{for } k = 2 \,, \quad P_{00} + P_{01} - P_{10} - P_{11} = \langle \hat{\cal V}_2 \rangle \,, \\
	&\text{for } k = 3 \,, \quad P_{00} + P_{01} - P_{10} - P_{11} = \langle \hat{\cal V}_3 \rangle \,, \\
	&\text{for } k = 4 \,, \quad P_{00} - P_{01} + P_{10} - P_{11} = \langle \hat{\cal V}_4 \rangle \,, \\
	&\text{for } k = 5 \,, \quad P_{00} - P_{01} - P_{10} + P_{11} = \langle \hat{\cal V}_5 \rangle \,, \\
	&\text{for } k = 6 \,, \quad P_{00} - P_{01} - P_{10} + P_{11} = \langle \hat{\cal V}_6 \rangle \,, \\
	&\text{for } k = 7 \,, \quad P_{00} - P_{01} + P_{10} - P_{11} = \langle \hat{\cal V}_7 \rangle \,.
\end{align*}

Each expectation value $\langle \hat{\cal{V}}_k \rangle$, for $k = 1, \ldots, 7$, was obtained by performing 8000 two-qubit independent measurements on the $ibmq_{-}osaka$ machine \cite{IBMQ}. The expectation value results from the experiments are $\langle \hat{\cal V}_1 \rangle \simeq 0.625005$, $\langle \hat{\cal V}_2\rangle \simeq -0.313760$, $\langle \hat{\cal V}_3 \rangle \simeq 0.654026$, $\langle \hat{\cal V}_4 \rangle \simeq 0.316057$, $\langle \hat{\cal V}_5 \rangle \simeq -0.646405$, $\langle \hat{\cal V}_6 \rangle \simeq 0.659897$, and $\langle \hat{\cal V}_7 \rangle \simeq 0.342302$. Thus, in view of Eq.~\eqref{J_2_in_terms_of_Vk}, the experimentally obtained value of probability current in the $N = 2$ case is
\begin{equation*}
	J = {1 \over 4 \pi} \langle \hat{\cal J}_2 \rangle \simeq -0.102789 \,.
\end{equation*}
The experimental value is negative, signifying QB, and is very close to the theoretical one, given by Eq.~\eqref{J_theor_N_2}. The relative error is approximately $3.1\%$.

\section{Conclusion}
\label{sec:conclusion}

In this paper, we report a quantum simulation of QB within a circular geometry. The system under consideration -- a particle moving freely in a circular ring -- is well-suited for the simulation, as its discrete angular momentum eigenstates can be effectively modeled using qubits. More specifically, $N$-qubit circuits can be employed to simulate QB for quantum states that are superpositions of $2^N$ angular momentum eigenstates. We explicitly design such circuits for the cases when $N = 1$ and $N = 2$, and subsequently implement them on the IBM-Q quantum computer. Our quantum simulations demonstrate negative probability current for quantum states comprised solely of non-negative angular momentum states, thereby confirming the presence of QB. The simulated probability current values are reasonably close to the corresponding theoretical values.

The quantum simulations presented in this paper have thus far been limited to one and two qubits, i.e., to the cases of $N = 1$ and $N = 2$, due to us having access only to noisy quantum devices. Nonetheless, we have established a comprehensive theoretical framework for simulating QB with an arbitrary number of qubits (i.e., for any value of $N$). To elaborate, a quantum simulation of QB involves two main stages: (i) the preparation of a backflowing state, and (ii) the subsequent measurement of probability current. Concerning (i), we have devised a specific example of a backflowing state valid for arbitrary $N$; this example is detailed by Eqs.~\eqref{state} and \eqref{example_state}. As for (ii), we have derived a universally applicable decomposition of the probability current operator, given by Eq.~\eqref{decomposition_of_J_N}, which can be readily employed to construct a current-measuring circuit for arbitrary $N$. In light of these findings, our theoretical results provide a clear pathway for simulating QB using any number of qubits.

The exploration of novel applications for quantum computing has been an active area of research \cite{GKZ+10Quantum,Som15,Bau22}. The study reported in the present paper expands the domain of potential applications of quantum computing and uncovers new connections between different areas of quantum physics.

\begin{acknowledgments}
	JJ acknowledges the support from the Basic Science Research Program through the National Research Foundation of Korea (NRF) funded by the Ministry of Education, Science and Technology (NRF-2021M3H3A1038085) and the Institute for Information \& communications Technology Promotion (IITP) grant funded by the Korea government(MSIP) (No.~2019-0-00003, Research and Development of Core technologies for Programming, Running, Implementing and Validating of Fault-Tolerant Quantum Computing System).

	We acknowledge the utilization of IBM Quantum services in conducting this work. The views expressed are those of the authors and do not reflect the official policy or position of IBM or the IBM Quantum team.
\end{acknowledgments}

%

\end{document}